\documentstyle[11pt,aasms,tighten,flushrt]{article}
\def \AU{~\rm{AU}}
\def \kpc{~\rm{kpc}}
\def \lesssim{\mathrel{<\kern-1.0em\lower0.9ex\hbox{$\sim$}}}
\def \gtrsim{\mathrel{>\kern-1.0em\lower0.9ex\hbox{$\sim$}}}

\begin{document}


\title{SPHERICAL PLANETARY NEBULAE}

\author{ 
Noam Soker
\affil{
Department of Physics, University of Haifa at Oranim\\
Oranim, Tivon 36006, ISRAEL \\
soker@physics.technion.ac.il }}


\bigskip
\centerline {\bf ABSTRACT}

By examining their mass loss history and their distribution
in the galaxy I argue that spherical planetary nebulae (PNe)
form a special group among all planetary nebulae.
 The smooth surface brightness of most spherical PNe suggests
that their progenitors did not go through a final intensive
wind (FIW, also termed superwind) phase.
  While $\sim 70 \%$ of the PNe of all other PNe groups are closer
to the galactic center than the sun is, only $\sim 30 \%$ of
spherical PNe are; $\sim 70 \%$ of them are farther away from the
galactic center.
 These, plus the well known high scale height above
the galactic plane of spherical PNe, suggest that the progenitors of
spherical PNe are low mass stars having low metallicity.
 Although many stars have these properties, only $\sim 10 \%$
of all PNe are spherical.
 By comparing the galactic distribution of spherical PNe
to the metallicity evolution in the galaxy, I find that the
critical metallicity above which no spherical PNe are formed is
${\rm [Fe/H]} \sim -0.4$.
 I explain this as well as other properties of spherical PNe
in the context of the companion model for shaping PNe,
arguing that spherical PNe are formed from stars which had no
close companion, stellar or substellar, orbiting them.
 I discuss the connection of the proposed scenario to the recent
finding of extrasolar planets and to the presence of
blue horizontal branch stars in globular clusters.




\section{INTRODUCTION}

 Most planetary nebulae (PNe) have a large scale axisymmetrical
rather than spherical structure.
 Axisymmetrical PNe can be classified into two main groups;
bipolar PNe, which are defined as PNe whose main structure contains
two lobes with an equatorial waist between them, and elliptical PNe,
which have a general elliptical shape, but no lobes, or only small ones.
 In the present paper I refer to PNe having a large and dense
concentration of mass in their equatorial plane, e.g., a ring,
but without lobes, as extreme elliptical PNe.
 Whereas bipolar and elliptical PNe were the focus of many
studies in recent years, basic questions regarding spherical PNe
were not discussed much in the literature.
 The only point mentioned in some papers is a large scale height
above (or below) the galactic plane of spherical PNe distribution,
which implies that they are formed from low mass stars (e.g.,
Manchado {\it et al.} 2000).
 The question most often addressed is:
what is the mechanism for forming non-spherical, i.e., bipolar or
elliptical, structures?
 In the present paper I reverse the emphasis by asking: 
{\it what is required to form a spherical PN?}
 Presenting the question this way allows me to consider
a wider scope, e.g., the formation of blue horizontal
branch (HB) stars.
 The two type of objects, I suggest, are connected in that both
 spherical PNe and blue HB stars are formed from low mass stars,
but blue HB stars will not form PNe.
 Therefore, the process behind the formation of blue HB stars
will prevent a star from forming a PN.

 Manchado {\it et al.} (2000) argue that $\sim 25 \%$ of the PNe in
their sample are round.
 In the present paper I use the term spherical PNe, assuming that
the PNe I classified as spherical are indeed spherical and not 
pole-on elliptical or bipolar PNe.
  I use my earlier classification of spherical PNe (Soker 1997,
hereafter S97), where more stringent criteria were used, and
where only 18 PNe are classified as spherical (Table 2 of S97).
These amount to $4 \%$ of the 458 PNe classified there,
but the estimate is that $\sim 10\%$ of all PNe are spherical.
 Manchado {\it et al.} (1996), for example, classify LSA 1
(PNG 029.8-07.8), K3-73 (PNG 084.0+09.5), and A 33
(PNG 238.0+34.8) as round PNe, but I can clearly see two arcs within
each of these PNe, which suggest that they are elliptical PNe.
 On the other hand, K1-14 (PNG 045.6+24.3) and A71
(PNG 084.9+04.4), which are classified as round PNe by
Manchado {\it et al.} (1996), but not by S97,
may have been spherical PNe before distortion.
 Another interesting case is A30 (PNG 208.5+33.2), which was
classified as round by Manchado {\it et al.}, but not by S97.
 The reason for the S97 classification of A30 as elliptical PN is the
knots close to its central star, which show an axisymmetrical
structure. But beside these four knots the nebula shows
all the  properties of a spherical PN.
 
 In section 2 some aspects of the distribution of these spherical
PNe in the galaxy and their mass loss history are studied.
 In section 3 these properties are related to the metallicity of
 their progenitors.
 In section 3 I discuss also the formation of spherical PNe in
 relation to blue HB stars and to the more than 50 known extrasolar
 planets, arguing that the findings are compatible with the
planet-induced axisymmetrical mass loss model.
 In this model low departure from spherical mass loss of some
stars on the upper AGB are attributed to planets
(Soker 2001a and references therein).
 Summary of main results, and some speculations and predictions, are
given in section 4. 
  
\section{PROPERTIES OF SPHERICAL PLANETARY NEBULAE}
\subsection{The Galactic Distribution}

 It is well known that spherical PNe have a larger average
galactic latitude $b$ than that of the other PNe (e.g., Manchado
{\it et al.} 2000).
 I now show that the class of spherical PNe, as classified by S97,
has also a larger galactic longitude than the other PNe.
 In S97 I classified the PNe according to the type of binary
 interaction that may have shaped them:
($i$) {\bf Spherical PNe,} which I assumed in S97 to be PNe whose
progenitors did not interact with any close companion, stellar or
substellar.
($ii$) {\bf Bipolar PNe.} Most, but not all, bipolar PNe
belong to this class. In S97 I assumed that these are PNe whose
progenitors interacted with close stellar companions
outside their envelopes.
($iii$) {\bf Extreme-Elliptical PNe.} This class includes some bipolar
PNe, but mainly extreme-elliptical PNe, e.g., having a ring but not
two lobes.
 I assumed in S97 that the progenitors of most of these
PNe interacted with a stellar companion via a common envelope phase.
($iv$) {\bf Elliptical PNe.} By elliptical alone, I refer to moderate
elliptical PNe which show only small departure from sphericity.
In S97 the progenitors of these PNe were assumed to have interacted with
substellar objects, but not with a close stellar object.
 In a later paper (Soker 2001b) I argue that many PNe which belong
to classes $(iii)$ and $(iv)$ can be formed from the influence of
a companion outside the progenitor's envelope, if the companion 
is not too close and/or two massive, but still accretes and blows
its own collimated fast wind (CFW).

 In Table 1 I compare the galactic distribution of the
four classes listed by S97.
 The quantities that are given for each class are as follows.
 The total number of PNe in the class, $N_T$, the average galactic
longitude of all PNe in the class $l_a$, and the number of PNe
toward the galactic center,  i.e., having
$l<90^\circ$ or $l>270^\circ$, $N_{\rm in}$.
Also given are quantities just for PNe toward the
galactic center:  $l_{\rm in}$ is the average value of the galactic
longitude, $fb_{\rm in}$ is the fraction of PNe having galactic
latitude of $\vert b \vert >10^{\circ}$, and
$<b_{\rm in}>$ is the median value of $\vert b \vert$.
 $N_{\rm out}$ is the number of PNe away from the galactic center,
i.e., having $90^\circ < l < 270^\circ$. 
 Other quantities with subscript `out' have the same meaning
as those with subscript `in', but only for PNe away from the
galactic center.
 For statistics involving the galactic longitude I took the galactic
longitude of PNe with $180^\circ < l < 360^\circ$ to be
$l_s=360^{\circ}-l$. 
 
 One well known property and two new ones clearly emerge from
the Table.
 (1) As is well known (see, e.g., reviews by Corradi 2000 and
Manchado {\it et al.} 2000) bipolar PNe are concentrated toward the
galactic plane, elliptical have intermediate scale height, while
spherical have the largest scale height.
The relevant quantities in the Table which show this behavior
are $fb_{\rm in, out}$ and $<b_{\rm in, out}>$.
 (2) The number of PNe toward and away from the galactic center
in each class, $N_{\rm in}$ and $N_{\rm out}$, and the average galactic
latitude $l_a$, reveal that spherical PNe tend to be concentrated
{\it away} from the galactic center.
While the differences in the fraction of PNe away from the
galactic center, $N_{\rm out}/N_T$, between the three classes
of nonspherical PNe are within the statistical errors,
the value for spherical PNe is more than twice as large. 
 (3) Another new property is the relatively very high average
galactic latitude of spherical PNe toward the galactic center.
  This is seen from comparing the values of $fb_{\rm in}$
and $<b_{\rm in}>$ with $fb_{\rm out}$ and $<b_{\rm out}>$.
For the three classes of nonspherical PNe  
$fb_{\rm in} < fb_{\rm out}$ and $<b_{\rm in}> < < b_{\rm out}>$,
while for spherical PNe opposite inequalities hold.
 Another way to see this is by noticing that there is a smooth
increase in $fb_{\rm out}$ and $<b_{\rm out}>$ when moving from bipolar
to extreme-elliptical to elliptical and to spherical, while
in $fb_{\rm in}$ and $<b_{\rm in}>$ the smooth increase holds
only for the nonspherical PNe, with a sharp rise between
elliptical and spherical PNe. 
 This behavior is less certain because of the poor statistics,
as there are only six spherical PNe toward the galactic center.
Also, the relevant physical  quantity is the galactic height,
for which the distance to each PN has to be known.
 I use the galactic latitude in the statistical analysis in order
to avoid the large uncertainties in distances to  PNe. 
 Despite these drawbacks, in the next section I suggest that this
behavior is real, and try to explain it.

 As discussed in the previous section, Manchado {\it et al.} (1996, 2000)
use a different classification for spherical PNe.
 In their classification properties (2) and (3) listed above
are not found.
 As also mention in the previous section, I disagree with Manchado
{\it et al.} (1996) in the classification of many PNe as spherical .
 Further study is clearly needed here.

\subsection{Mass Loss History}

 From the structures of most PNe, as well as other considerations,
we know that most stars terminate the AGB by blowing a superwind
(Renzini 1981), i.e., a final intensive wind (FIW), which is not
faster than the regular AGB wind but has a much higher mass loss rate.
 In PNe this wind forms a dense shell.
 In many cases, inward to this shell there is a bright rim formed
by the fast wind from the central star, and a fainter shell or
halo outside the shell (Frank, Balick, \& Riley 1990).
 Examining the images of the 18 spherical PNe (S97 Table 2),
I find that most do not show signatures of a FIW (Soker 2000a).
I find dense shells only in Bd+30 3639 and IC 3568, both
which were suspected of being elliptical seen pole-on
(marked PO by S97), H3-75, and possibly in A 15 and Lo 4.
  As discussed in Soker (2000a), where more details and references
can be found,  the same correlation holds in many
elliptical PNe; their outer faint halo is spherical, or only
slightly elliptical, while the dense shell, which was formed from
the FIW (superwind), is highly nonspherical.
 The positive correlation, albeit not perfect, between the onset
of a FIW and a more nonspherical mass loss geometry can
in principle result from two types of mechanism (Soker 2000a).
In the first type ($\S 2$ of Soker 2000a) a primary process or event
causes both the increase in the mass loss rate and its deviation from
spherical geometry.
 A primary mechanism or event may be external, e.g., a late
interaction with a stellar or substellar companion, or internal,
e.g., rapid changes in some of the envelope properties.
 In the second type of flow, the FIW (superwind) makes possible
a mechanism which is very inefficient at low mass loss rates.
 For example, the high mass loss rate leads to an optically thick
flow, where inner regions shield outer regions from the stellar radiation,
allowing enhanced dust formation rate.
 Such a process above cool magnetic spots will lead to the formation
of an elliptical nebula if the spots are concentrated
around the equator (Soker 2000b).
 
 Finding elliptical faint (low density) halos will indicate that
a nonspherical mass loss can occur at low mass loss rates.
 It is not clear yet if such halos exist. 
 Finding spherical PNe which show signatures of superwind would
indicate that the mechanism behind the FIW does not necessarily
lead to axisymmetrical mass loss.
 There are not many such PNe, and the few mentioned above may be
elliptical PNe seen edge-on.
 The explanation for the positive correlation between mass loss rate
and mass loss geometry may therefore require a more elaborate answer.
 In many elliptical PNe, however, the FIW (superwind) is present
along the polar directions as well as near the equatorial plane,
although mass loss rate is higher near the equatorial plane.
 This seems to indicate that the second mechanism is correct:
the mass loss rate increases at all directions, but it allows further
increase of the mass loss rate near the equatorial plane,
probably only if some minimum value of stellar rotation exists.
I argue that this minimum rotation rate requires the AGB envelope
to be spun-up by a companion (Soker 2001a), stellar or substellar.
 If this is the case, it is quite possible that some round PNe
showing signatures of FIW will indeed be spherical, and not
pole-on elliptical PNe.

\section{WHAT IS REQUIRED FOR BECOMING A SPHERICAL PLANETARY NEBULA?}

\subsection{The Metallicity Connection}

 I propose the following explanation for the findings of the
previous section.
 To form a spherical PNe the progenitor should have
a metallicity below a critical value.
 For a given mass, the probability of forming a spherical PN
increases as metallicity decreases, starting from zero
probability at the critical metallicity 
${\rm [Fe/H]_c}$.
 The star should also be above some lower mass limit.
For a given metallicity, the probability of forming a
spherical PN increases with the progenitor initial mass $M_i$.
 I further argue, as I have been doing for the last decade,
that this is expected if planets play a significant role in
spinning-up evolved stars (Soker 2001a and references therein).
 To show that this is compatible with the galactic distribution
of spherical PNe, I consider the galactic-metallicity dependence
on age and the distance to the galactic center.
 Hereafter the age $t$ is expressed in Gyr (being positive with
$t=0$ at present), while the distance to the galactic center,
the Galactocentric radius, will be with respect to the distance
of the sun to the galactic center in units of kpc ($r=0$ at the
solar neighborhood, and $r>0$ moving away from the galactic center).
 
 Lineweaver (2001) studies the relation between the metallicity
of sun-like stars and the presence of hot Jupiter-like planets
orbiting them. 
 Lineweaver uses a galactic metallicity evolution in the solar
neighborhood which presently has ${\rm [Fe/H]_s} = -0.2$,
similar to the metallicity given by Allen, Carigi, \& Peimbert (1998)
in their PNe study, but lower than values given by others, e.g.,
Carraro, Ng \& Portinari (1998; see reviews by Henry \& Worthey 1999
and Shields 2001).
 I take a metallicity evolution in the solar neighborhood
to be some average of these studies, 
\begin{eqnarray}
{\rm [Fe/H]_s}(t) =-0.1-0.02t-a(t-5)^4 \qquad
\left\{\begin{array}{ll} a=0 & \mbox{ $t \leq 5$} \\
         a= 5 \times 10^{-4} & \mbox{ $5 < t \lesssim 10$,}
                         \end{array} \right. 
\end{eqnarray}
where $t$ is the age given in Gyr,
and subscript $s$ refers to the solar neighborhood.
 Allen {\it et al.} (1998) deduce from their PNe study that the
metallicity gradient, with respect to the distance from the galactic
center, was steeper in the past, while Carraro  {\it at al.} (1998)
find a nonmonotonic evolution of the metallicity gradient which, they
argue, is compatible with a gradient that does not change with time.
 Again, I take the gradient evolution to be the average between
these two studies,
\begin{eqnarray}
\frac {d {\rm [Fe/H]}}{dr} = 
-2\times 10^{-4} t^2 - 0.001 t -0.09 ~{\rm dex} \kpc^{-1},
\end{eqnarray}
 where $r$ is the distance from the galactic center minus
the solar distance to the galactic center (in kpc).
 The metallicity is therefore given by
\begin{eqnarray}
{\rm [Fe/H]}(t,r)={\rm [Fe/H]}_s+\frac {d {\rm [Fe/H]}}{dr}r.
\end{eqnarray}
 The contour map of ${\rm [Fe/H]}(t,r)$ is plotted in Figure 1.
 The contour levels have a constant spacing of $0.1 {\rm dex}$,
with values marked near each contour line.

 For present purposes the exact values of the metallicity are
not important, but only the general variation of metallicity with
$r$ and $t$.
 To demonstrate the proposed explanation for the spherical PNe
properties, I estimate the critical metallicity to be
${\rm [Fe/H]_c} \simeq -0.4$, and mark this contour by a
thicker line.
 If only stars with ${\rm [Fe/H]}<{\rm [Fe/H]_c} \simeq -0.4$ can
form spherical PNe, it is clear from the figure that more
spherical PNe will be located at larger distances from the
galactic center, i.e., $r>0$, and they all will be descendants
of low mass progenitors.
 These PNe occupy the upper left corner of Figure 1, where the
large age means that these PNe are formed from low mass stars.
 These low mass star progenitors imply that the galactic distribution
of the descendant PNe will have a large average galactic
latitude, as is indeed observed (Manchado {\it et al.} 2000).
 Moreover, the spherical PNe closer to the galactic center, $r<0$,
with ${\rm [Fe/H]} \gtrsim -0.4$, are on average lighter (older) than those
farther away from the galactic center, $r>0$, meaning a larger
average galactic latitude.
 This is compatible with the findings of section 2 (compare 
$fb_{\rm in}$ and $<b_{\rm in}>$ with $fb_{\rm out}$ and $<b_{\rm out}>$,
all given in Table 1).

 To have $\sim 5-10 \%$ spherical PNe among all PNe, the probability for
forming a spherical PNe must increase substantially as metallicity
decreases by $\sim 0.3-0.4~ {\rm {dex}}$ from the critical value. 
 In the case assumed here, it is from ${\rm [Fe/H]_c} \simeq -0.4$ to
${\rm [Fe/H]} \simeq -0.8$.
 Two other relevant phenomena also change significantly when metallicity
is changed by $\sim 0.3-0.4 {\rm dex}$.
 The first is the presence of hot-Jupiter stars around sun-like stars.
 The several tens of extrasolar planets that have been found
tend to orbit metal-rich stars.
  Lineweaver (2001) analyzes 32 hosts of Jupiter-like planets, and
argues that the probability of a sun-like star hosting a hot
Jupiter-like planet increases from $\lesssim 10 \%$ at
${\rm [Fe/H]} = 0.1$ to $\sim 90 \%$ at ${\rm [Fe/H]}=0.4$,
and to almost $100 \%$ at ${\rm [Fe/H]}=0.6$.
 The second phenomenon is the distribution of HB
stars on the Hertzsprung-Russel diagram of globular clusters
(the HB morphology).
The distribution, e.g., the relative number of blue HB
stars, varies from one globular cluster to another.
 It has long been known that metallicity is the main, but not sole,
factor which determines the HB morphology
(for a historical review see, e.g., Rood, Whitney, \& D'Cruz 1997;
Fusi Pecci \& Bellazzini 1997).
 The other factor(s) which determine(s) the HB morphology
is termed the `second parameter', and it is commonly thought that
it has to do with mass loss on the red giant branch (Rood 1973).
 It is well known that the HB morphology
significantly changes its behavior in a relatively narrow
metallicity range of $-1.75 \lesssim {\rm [Fe/H]} \lesssim -1.3$
(Soker \& Hadar 2001 and references therein).
 Therefore, it is not unlikely that the property which determines
the probability of forming a spherical PN significantly changes
over a narrow metallicity range, around ${\rm [Fe/H]} \sim -0.5$.
 
\subsection{The Planets Conjecture}

 I now relate the metallicity dependence on time and Galactocentric
radius to the planet-induced axisymmetrical mass loss model.
 In this model low departure from spherical mass loss of some
stars on the upper AGB is attributed to planets, or brown dwarfs,
which spin-up the AGB stellar envelopes (Soker 2001a and references
therein).
 In other stars the nonspherical mass loss is due to interaction with
stellar companions.
 Therefore, if a massive enough and close enough planet is present
around a star, the stellar envelope will be spun-up by the
planet via tidal effects and common envelope evolution.
  The maximum orbital separation for an interaction to take place
is $\sim 4 \AU$.
  In a previous paper (Soker 2001a) I argue that even planets having
masses as small as $0.01 M_J$, where $M_J$ is Jupiter's mass, can lead to
a slightly nonspherical mass loss geometry.
 Low mass stars hosting a close planet will engulf the planet already
on the first giant branch (RGB), prior to the HB.
The stars will be spun-up, their mass loss rate is likely to increase
(Soker \& Hadar 2001), and they will retain very small mass in
their envelopes, forming blue HB stars.
 Therefore, it is very likely that most known extrasolar systems
will not form PN at all, but rather the hosting stars will engulf the
orbiting planets during their RGB, lose most of their envelope on the
RGB, and turn into blue HB stars.
 Blue HB stars are not likely to reach the upper AGB
and form PNe because of their very low envelope masses.
 Even single low mass stars may lose too much
mass on the RGB, possibly due to fast rotation, and never form PNe.
 For different reasons Allen {\it et al.} (1998) argue that
less than half of all stars of initial mass $M_i<1.3M_\odot$ form PNe.

 It seems therefore that stars hosting close planets with
mass $\gtrsim 0.01 M_J$ will either lose most of their envelope
on the RGB and never form PNe, or reach the AGB but form
nonspherical, i.e., elliptical, PNe.
 To form a spherical PN a star should not have any
close planet with a mass $\gtrsim 0.01 M_J$, and of course no close
stellar companion or brown dwarf either.
 It is very likely, e.g., Lineweaver (2001), that the probability of
forming a planet, and the planet's mass, strongly depend on 
metallicity.
 Lineweaver (2001), for example, assumes that Earth-type planets are
formed only when ${\rm [Fe/H]} \gtrsim -1$, with increasing probability
for increasing metallicity.
For planets to induce mass loss in globular clusters, low mass planets
should be formed already at ${\rm [Fe/H]} \sim -1.7$.
 The initial mass of globular clusters stars now reaching the
HB is $\sim 0.9$, so Earth-like planets maybe enough in some cases
to enhance mass loss. More massive planets may be required to prevent
stars of $ M_i \gtrsim 1.1 M_\odot$ from reaching the AGB.
 It is not unlikely that planets with mass of $\gtrsim 0.01-0.1 M_J$
will be formed at high probability when
${\rm [Fe/H]} \gtrsim -0.4$, as I suggested in the previous subsection.
 Such close planets will either cause their hosting star to form
an elliptical PN or cause the star to lose most of its envelope,
never forming a PN.

\section{SUMMARY}

 The main finding of the paper concerns the distribution of spherical
 PNe in the galaxy.
 I found that spherical PNe tend to be concentrated {\it away} from
the galactic center, and spherical PNe which are closer than the sun
to the galactic center have a relatively very high average
galactic latitude (Table 1).
 I used this finding to argue that spherical PNe are formed from
low metallicity stellar progenitors (see Figure 1).
 A crude estimate suggests that spherical PNe are formed from stars
with ${\rm [F_e/H]} \lesssim -0.4$ (and initial mass of
$M_i \gtrsim 0.9 M_\odot$), although only a minority of stars with these
properties do form spherical PNe. 

I further argued that the dependence on metallicity may result from
the role played by planets in spinning-up the envelope of RGB and AGB
stars.
 A planet will either spin-up the stellar envelope to blow
a nonspherical wind, hence forming an elliptical PN, or, if the
spin-up occurs on the RGB of a low mass star, the relatively fastly
rotating RGB star will lose most of its envelope and form a blue HB
star, but will never reach the upper AGB and form a PN.
 Lineweaver (2001) finds the probability of forming Jupiter-like planets,
with masses of $M_p \sim M_J$, to increase from $\lesssim 10 \%$
at ${\rm [Fe/H]_l} = 0.1$ to $\sim 90 \%$ at ${\rm [Fe/H]_u}=0.4$.
 The  planet-induced axisymmetrical mass loss model to explain the
properties of spherical PNe requires that the probability of forming
planets of $M_p \sim 0.01-0.1 M_J$ increase significantly from 
${\rm [Fe/H]_l} \simeq -0.8$ to  ${\rm [Fe/H]_u} \simeq -0.4$.
 Of course, more massive planets can be formed as well.
 For comparison, the planet second parameter model to explain some
HB morphologies of globular clusters (Soker \& Hadar 2001) requires 
that the probability of forming planets of $M_p \simeq 0.01 M_J$ increase
significantly from ${\rm [Fe/H]_l} \simeq -1.7$ to
${\rm [Fe/H]_u} \simeq -1.3$.
  These three critical metallicity bands show a monotonic trend,
which, I claim, supports the conjecture that planets play a crucial
role in the mass loss history of evolved stars. 

\acknowledgments
This research was supported in part by a grant from the 
US-Israel Binational Science Foundation.

\bigskip

{\bf FIGURE CAPTIONS}
{\bf Fig. 1.$-$} Metallicity contour map.
The value of [Fe/H] is written near each contour line,
$r$ is the Galactocentric distance relative to the sun
($r=0$ for the solar neighborhood, $r>0$ is away from the galactic
center) and $t$ is the age ($t=0$ at present).
 This contour map is used in the text together with the distribution
 of spherical PNe in the galaxy (Table 1) to argue that the
progenitors of spherical PNe have ${\rm [Fe/H]} \lesssim -0.4$
(marked by a thick line).

\end{document}